\newcommand\techrep

\ifdefined\techrep
\documentclass[english]{article}
\usepackage{fullpage}
\usepackage[colorlinks,naturalnames,citecolor=blue,urlcolor=blue]{hyperref}

\else
\documentclass[english]{llncs}
\usepackage[colorlinks,naturalnames,citecolor=blue,urlcolor=blue]{hyperref}
\fi

\newcommand{\delete}[1]{}
\usepackage{amsmath}

\usepackage{lmodern}

\usepackage[T1]{fontenc}
\usepackage[latin9]{inputenc}
\usepackage{float}
\usepackage{wasysym}

\makeatletter

\floatstyle{ruled}
\newfloat{algorithm}{tbp}{loa} 

\usepackage{tikz}
\usetikzlibrary{arrows,positioning,calc,fit}

\@ifundefined{showcaptionsetup}{}{%
 \PassOptionsToPackage{caption=false}{subfig}}
\usepackage{subfig}
\makeatother

\usepackage[english]{babel}

\newcommand{\red}[1]{{\color{red}{#1}}}
\newcommand{\hide}[1]{}

\global\long\def\real{\mathsf{real}}
\global\long\def\tunit{\mathsf{unit}}
\global\long\def\para{\,\|\,}
\global\long\def\num{\mathsf{numeral}}
\global\long\def\suc{\mathsf{Suc}}
\global\long\def\And{\mathsf{And}}
\global\long\def\Add{\mathsf{Add}}
\global\long\def\bool{\mathsf{bool}}

\global\long\def\ptto{\stackrel{\circ}{\rightarrow}}
\global\long\def\const{\mathsf{Const}}
\global\long\def\True{\mathsf{True}}
\global\long\def\False{\mathsf{False}}

\global\long\def\inte{\mathsf{int}}
\global\long\def\zero{\mathsf{zero}}
\global\long\def\Id{\mathsf{Id}}
\global\long\def\plus{\mathsf{plus}}
\global\long\def\mult{\mathsf{mult}}
\global\long\def\simu{\mathsf{simulink}}
\global\long\def\nat{\mathsf{nat}}
\global\long\def\tp{\,'\!}
\global\long\def\st{\,.\,}

\ifdefined\techrep
\newcommand{\rs}{}
\newcommand{\rd}{}
\else
\newcommand{\rs}{\vspace{-1ex}}
\newcommand{\rd}{\vspace{-0.4ex}}
\fi

\begin{document}

\ifdefined\techrep
\title{Type Inference of Simulink Hierarchical Block Diagrams in Isabelle\thanks{This work has been partially supported by the Academy of Finland and the U.S. National Science Foundation (awards \#1329759 and \#1139138).}}
\author{Viorel Preoteasa$^1$ \and Iulia Dragomir$^2$ \and Stavros Tripakis$^{1,3}$\\[1ex]
\small$^1$ Aalto University, Finland ~~~ $^2$ Verimag, France ~~~ $^3$ University of California, Berkeley, USA}

\else
\title{Type Inference of Simulink Hierarchical Block Diagrams in Isabelle\thanks{This work has been partially supported by the Academy of Finland and the U.S. National Science Foundation (awards \#1329759 and \#1139138).}}
\author{Viorel Preoteasa\inst{1} \and Iulia Dragomir\inst{2} \and Stavros Tripakis\inst{1,3}}
\institute{Aalto University, Finland \and Verimag, France \and University of California, Berkeley, USA}
\fi

\maketitle

\begin{abstract}
Simulink is a de-facto industrial standard for the design of
embedded systems. In previous work, we developed a compositional analysis framework 
for Simulink models in Isabelle -- the Refinement Calculus of Reactive Systems (RCRS), 
which allows checking compatibility and substitutability of components.
However, standard type checking was not considered in that work.
In this paper we present a method for the type inference
of hierarchical block diagrams using the Isabelle theorem prover.
A Simulink diagram is translated into an (RCRS) Isabelle theory.
Then the Isabelle's powerful type inference mechanism is used to infer the  types of the diagram based on the types of the basic blocks.
One of the aims is to handle formally as many diagrams as possible.
In particular, we want to be able to handle even those diagrams that may have typing ambiguities,
provided that they are accepted by Simulink.
This method is implemented
in our toolset that translates Simulink diagrams into Isabelle theories 
and simplifies them.
We evaluate our technique on several case studies, most notably, an
automotive fuel control system benchmark provided by Toyota.
\end{abstract}

\ifdefined\techrep \else \vspace{-3ex} \fi
\section{Introduction}
\label{sec:intro}

Simulink is a widespread
tool from Mathworks for modeling and simulating embedded control systems.
A plethora of formal verification tools exist for Simulink, both from
academia and industry, including Mathwork's own Design Verifier.
Formal verification is extremely important, particularly for safety critical
systems. Formal verification techniques make steady progress and are increasingly
gaining acceptance in the industry.

At the same time, we should not ignore more ``lightweight'' methods, which 
can also be very beneficial. In this paper, we are interested in particular
in type checking and type inference. Type checking is regularly used in many
programming languages, as part of compilation, and helps to catch many
programming mistakes and sometimes also serious design errors. 
Type inference is a more advanced technique which usually includes
type checking
but in addition permits types to be inferred when those are not given by the user,
thus automatically extracting valuable information about the design.
Importantly, both type checking and type inference are typically much less
expensive than formal verification.
We therefore view both of them as complementary to
formal verification for the rigorous design of safety-critical systems.

Simulink already provides some kind of type checking and inference as part 
of its basic functionality. In the version R2016b that we used while writing
this article, the user has to open a diagram and then click on {\em Display}
$\to$ {\em Signals \& Ports} $\to$ {\em Port Data Types}, upon which Simulink
(computes and) displays the typing information, for example, as shown in
Fig.~\ref{figsimulinkaccepts}.
Unfortunately, Simulink analyses are proprietary, and as such it is difficult
to know what type checking and inference algorithms are used.
Moreover, the way Simulink uses the typing information is often strange,
as illustrated by the examples that follow.

\ifdefined\techrep
\begin{figure}[t]
\centering
\includegraphics[width=.3\textwidth]{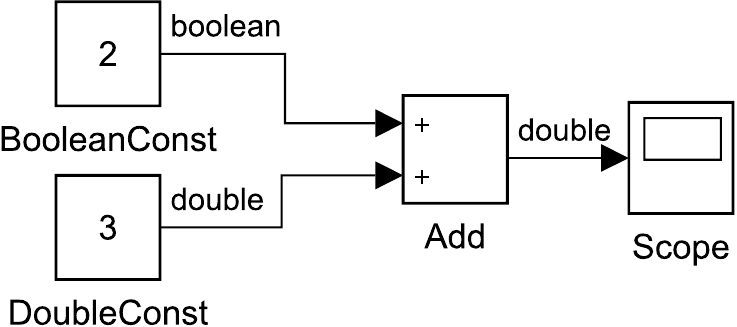} \qquad\qquad\qquad
\includegraphics[width=.4\textwidth]{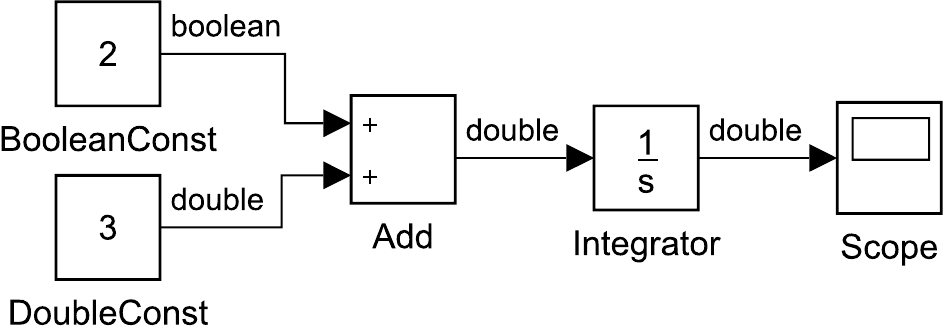}
\caption{Two Simulink diagrams. Both are accepted
(i.e., simulated) by Simulink.}
\label{figsimulinkaccepts}
\end{figure}
\else
\begin{figure}[t]
\centering
\includegraphics[width=.4\textwidth]{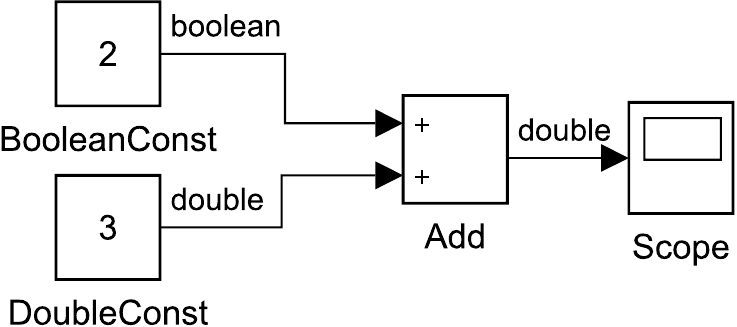} \qquad
\includegraphics[width=.5\textwidth]{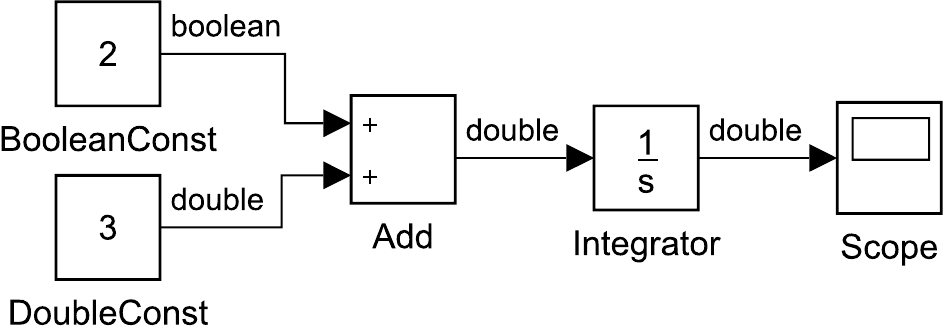}
\caption{Two Simulink diagrams. Both are accepted
(i.e., simulated) by Simulink.}
\label{figsimulinkaccepts} \vspace{-3ex}
\end{figure}
\fi

Consider first the two diagrams shown in Fig.~\ref{figsimulinkaccepts}.
Both these examples capture implicit type conversions performed by Simulink.
In both diagrams, there are two {\em Constant} blocks, with values $2$ and $3$
respectively. In the first block, we manually set the output type to be
{\em Boolean}. 
\ifdefined\techrep
This is done by clicking on the block and setting its {\em Output data type} under {\em Signal Attributes}.
\fi
In the second block
we manually set the output type to {\em double}. The outputs of the two
constants are fed into an {\em Add} block which performs addition. In the rightmost diagram, 
the result is fed into an {\em Integrator}. The block
{\em Scope} plots and displays the output over time.

Both diagrams of Fig.~\ref{figsimulinkaccepts} are {\em accepted} by
Simulink, meaning that they can be simulated. Although Simulink issues a
warning that says 
\hide{
\begin{quote}
``Parameter precision loss occurred for 'Value' of 'addBoolDouble/BooleanConst'. The parameter's value cannot be represented exactly using the run-time data type. A small quantization error has occurred.''
\end{quote}
}
``Parameter precision loss occurred \ldots A small quantization error has occurred.''
the results of the simulation appear as expected: a constant value $4$ in the
case of the leftmost diagram, and a straight slope from values $0$ to $40$ for
the rightmost diagram, when simulated from $0$ to $10$ time units.
Simulink performs an implicit conversion of
$2$ to the boolean value {\em true}, and then another implicit conversion of
{\em true} to the real value $1$, in order for the addition to be performed.
These implicit conversions are stipulated in the Simulink documentation 
(when the source block allows them).
Therefore, the result is $3+1=4$. 

Although these examples seem unsual, they are designed to be minimal
and expose possible problems, similar to those detected in a Fuel Control System (FCS) benchmark provided by Toyota~\cite{JinDKUB14}. 
It is common practice to mix, in languages that 
allow it, Boolean and numeric values in a way exposed by these examples. 
We have tested this behavior extensively and 
we have observed that other languages that perform automatic conversions
between Boolean and numeric values behave consistently with Simulink 
(e.g., C: 
$(\mathsf{double})3 + (\mathsf{bool})2 = 4.0$, Python: 
$\mathsf{float}(3) + \mathsf{bool}(2) = 4.0$).

\ifdefined\techrep
\begin{figure}[t]
\centering
\includegraphics[width=.3\textwidth]{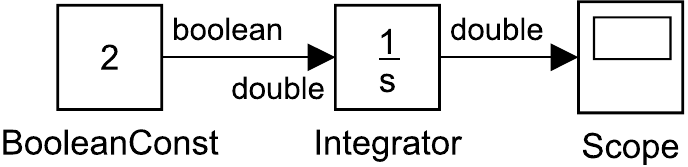}
\caption{A diagram rejected by Simulink.}
\label{figsimulinkrejects} 
\end{figure}
\else
\begin{figure}[t]
\centering
\includegraphics[width=.4\textwidth]{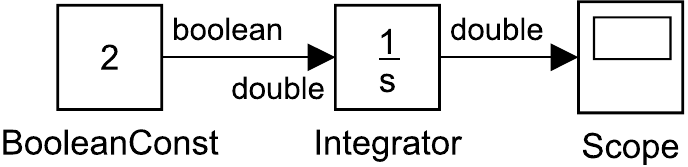}
\caption{A diagram rejected by Simulink.}
\label{figsimulinkrejects} \vspace{-3ex}
\end{figure}
\fi

Now, consider the diagram shown in Fig.~\ref{figsimulinkrejects}, where the
output of the same Boolean constant block as the one used in the previous
diagrams is fed directly into the integrator. 
In this case, Simulink {\em rejects} this diagram (meaning it refuses to
simulate it). It issues an error message saying:
``Data type mismatch. Input of Integrator expects a signal of data type 'double'. However, it is driven by a signal of data type 'boolean'.''
The Integrator, as well as other block types, 
accepts only inputs of type double and implicit conversions 
(from Boolean to double or vice-versa) are not allowed and performed. 
We remark that Simulink does not treat diagrams in a consistent way
with respect to typing. One of the goals of this paper is to present a formal type checking and 
inference framework for Simulink, where such examples are treated consistently
(and meaningfully).

\hide{
\begin{quote}
``Data type mismatch. Input port 1 of 'boolIntegrator/Integrator' expects a signal of data type 'double'. However, it is driven by a signal of data type 'boolean'.''
\end{quote}
}

\hide{
This choice appears arbitrary, as one might have expected Simulink to perform
an implicit conversion from $2$ to the boolean {\em true} and then to the real
number $1$, as in the case of the diagrams of Fig.~\ref{figsimulinkaccepts}.
One of the goals in this paper is to present a formal type checking and 
inference framework for Simulink, where such examples are treated consistently
(and meaningfully). 
}

The contribution of this work is a type inference mechanism for Simulink diagrams, on top of the type inference mechanism of the Isabelle theorem prover \cite{Nipkow:2002:IPA:1791547}. 
One important feature of this approach is handling Simulink basic blocks 
locally, without knowledge of their environment. The challenge
of this work is embedding the more relaxed type system of Simulink into the formal type system of Isabelle, 
while preserving the semantics, and as much typing information as possible. 
We apply this technique to several case
studies, including the FCS benchmark.

This work is part of a larger project on translating Simulink diagrams into
Isabelle theories suitable for 
analysis and verification. Because 
Isabelle's language is formal and precise, we can directly obtain concise and correct code in
other languages that can be used for processing Simulink models.
For example from the Isabelle model we easily obtain Python code for simulations,
and Z3 SMT solver \cite{DeMouraB08} model for automatically checking properties.

\hide{
Are there any diagrams that are \_not\_ accepted by Simulink?
\begin{itemize}
\item There are some diagrams that are not accepted by Simulink. For example
two constants that enter a conjunction, and the result enters an integral.
Simulink gives an error for this example. On the other hand, if in
between the conjunction and the integral we add an addition block
of the result of the conjunction and some constant, then Simulink
accepts it. Our translation, using the generic option creates meaningful
theories in both cases. 
\end{itemize}
}

\section{Related Work}
\label{sec:rwork}

The verification of Simulink diagrams has been extensively studied
in the literature, by proposing model transformations of Simulink
diagrams to a formal framework. Formal frameworks include Hybrid Automata \cite{AgrawalSK04}, 
BIP \cite{SfyrlaTSBS10}, NuSMV \cite{MeenakshiBR06}, Boogie \cite{ReicherdtG2014}, Timed Interval Calculus \cite{ChenDS09}, 
Function Blocks \cite{YangV12}, I/O Extended Finite Automata \cite{ZhouK12}, Hybrid CSP \cite{ZouZWFQ13}, and SpaceEx \cite{MinopoliF16}. 
Many of the target formalisms define a typing feature, and the proposed model translations make use of it: a basic block is mapped to some 
``expression'' on inputs and outputs, where the types of inputs and outputs are dependent of the block type. The static type checking is then 
delegated to the target framework, if such functionality is available. However, these studies mostly aim for formal verification of Simulink 
diagrams and do not report about type checking. Also, it is unclear to what extent the model transformations perform an analysis of 
inputs/outputs dimensions (scalar, vector, and matrix) to be used in the generated expressions.

The most relevant work with respect to type checking Simulink diagrams is described in \cite{TripakisSCC05} and \cite{RoyS11}. 
In \cite{TripakisSCC05}, a model transformation from discrete-time Simulink models to Lustre is presented, where type inference is used for computing the inputs/outputs type. 
Then the Lustre analyser performs type checking on the obtained model. In \cite{RoyS11}, the SimCheck contract framework for type checking the mathematical aspects of Simulink blocks is described. 
SimCheck allows the user to annotate ports and wires with types, but also with their units (e.g., \textit{cm}). A translation to Yices \cite{SRI:yices:tool} 
supports the automated checking of component compatibility within the model with respect to static and behavioral types. In contrast to SimCheck, we automatically infer the types and dimensions of 
signals from the Simulink diagrams, but we do not infer or check for physical units. 

In previous work, we have presented the {\em Refinement Calculus of Reactive Systems} (RCRS)~\cite{preoteasa:tripakis:2014,DBLP:conf/spin/DragomirPT16}, 
a compositional framework for static analysis of hierarchical block diagrams in general, and Simulink models in particular.
In the RCRS framework blocks are specified syntactically by general formulas
({\em contracts}) on input, output, and state variables. These contracts are
then composed using serial, parallel and feedback composition operators.
Such contracts can be seen as richer types, and the compatibility and contract
synthesis methods developed in RCRS can be seen as type checking and type
inference techniques. However, the contracts considered in RCRS are much more
powerful than the types considered in this paper, and the compatibility and
synthesis algorithms of RCRS are much more expensive (requiring in general
quantifier elimination and satisfiability checking in expressive logics).
Therefore, the framework proposed in this paper is much more lightweight.

\hide{
\red{Perhaps salvage some text from below? e.g., the remarks about PVS?\\
This also raises a question in my mind, and potentially also to reviewers:
we are selling RCRS as a \textquotedbl{}behavioral type checking/inference\textquotedbl{}
framework essentially, where the types are the contracts / relational
interfaces / predicate transformers. So why do we need yet another
type framework? 
\begin{itemize}
\item I think that this does not prevent us to use standard type checking
techniques that are decidable for distinguishing between Booleans
and reals, and use our technique for more advanced features such as
detecting division by zero.
\item PVS for example, although has subtypes, and dependent types, can also
automatically decide that an expression is real and not Boolean, although
it may not be able to decide automatically that the expression is
a nonnegative real.
\end{itemize}
}

\red{we probably also need to say something about type theory, although NFM might not contain many programming language folks. but in any case, one might ask "what is new here from a type theory perspective?}
}

In this work we use the Isabelle theorem prover which has 
a standard type inference mechanism.
Our goal is to give an embedding of Simulink into a language and framework suitable for further processing (simplifications, checking of properties, and even simulation). 
Other systems for logical reasoning (PVS \cite{OwreRS92}, Z3, \ldots) could also be used
for this purpose. Some of the problems encountered in our translation  may have simpler solutions in these systems, but some other 
aspects may become more difficult. All translation attempts to 
languages that do not allow automatic conversions between Boolean and numeric  values would have similar challenges as the ones that we solve
here.

\section{Preliminaries}
\label{sec:prelim}

\subsection{Isabelle}
\label{subsec:isabelle}

Isabelle/HOL is an interactive theorem prover based on higher order
logic. Isabelle provides an environment which consists of a powerful
specification and proving language and it has a rich theory library
of formally verified mathematics. Notable features of Isabelle include
a type system with type inference, polymorphism and overloading, and
axiomatic type classes.

Isabelle's type system includes the basic types $\bool,\real,\inte,\nat$,
type variables $\tp a,\tp b$, etc., and predefined type constructors $\tp a\to\tp b$
(functions from $\tp a$ to $\tp b$) and $\tp a\times\tp b$ (Cartesian
product of $\tp a$ and $\tp b$). For term $f(x,g(y))$ we can
specify that it has a type $t$ by using $:t$ after the term $f(x,g(y)):t$.
Isabelle uses type inference to deduce the type of a term. For example
the next declaration introduces a function $f:\tp a\to\tp b\to(\tp a\to\tp b\to\tp c)\to\tp c$
\rs
$$
\mathsf{definition\ }f(x)(y)(g)=g(x)(y) \rd
$$
The type of $f$ is the \emph{most general type} such that the expression
$f(x)(y)(g)=g(x)(y)$ is \emph{well typed}. 
A type $t$ is {\em more general} than a type $t'$ if $t'$ can be
obtained from $t$ by instantiating the type variables in $t$ with
some {\em type expressions}.

\ifdefined\techrep
We can also use \emph{lambda
abstraction} to define $f$:
\rs
\[
\mathsf{definition\ }f=(\lambda x,y,g\st g(x)(y)) 
\]
The term $f(\lambda x.g(x))(y)$ has the type $((\tp a\to\tp b)\to\tp c\to\tp d)\to\tp d$.
\fi
We can also use specific types in definitions:
\rs
\[
\mathsf{definition\ }h(x:\real)(y)(g)=g(x)(y) \rd
\]
In our translation of Simulink to Isabelle we use the type inference
mechanism.

Another important feature of Isabelle that we use is the type classes \cite{Haftmann2007}.
This is a mechanism that can be used, for example, to overload a polymorphic
function $+:\tp a\to\tp a\to\tp a$ on different types for $\tp a$.
\rs
\ifdefined\techrep
\[
\begin{array}{l}
\mathsf{class\ plus=} \\
\quad\mathsf{fixes}\ +:\tp a\to\tp a\to\tp a \\[1ex]
\mathsf{instantiation\ nat:plus}\\
\quad\mathsf{definition\ }0+x=x \  |  \ \suc(x)+y=\suc(x+y)\\[1ex]
\mathsf{instantiation\ real:plus}\\
\quad\mathsf{definition\ }x+y=...
\end{array} 
\] 
\else
\[
\begin{array}{ll}
\mathsf{class\ plus=} & \mathsf{instantiation\ real:plus}\\
\quad\mathsf{fixes}\ +:\tp a\to\tp a\to\tp a &
\quad\mathsf{definition\ }x+y=\ldots
\\[1ex]
\mathsf{instantiation\ nat:plus}\\
\quad\mathsf{definition\ }0+x=x \  |  \ \suc(x)+y=\suc(x+y)\\[1ex]
\end{array} \rd
\] 
\fi
We define the type class plus with the constant $+$ of polymorphic
type $\tp a\to\tp a\to\tp a$, and two instantiations to natural and
real numbers. In a term $x+y$, the type of $x$ and $y$ is not
just a type variable $\tp a$, but a type variable $\tp a$ of class
$\plus$. This is represented syntactically as $x:\tp a:\plus$. The
terms $(x:\nat)+y$ and $(x:\real)+y$ are well typed because the
types $\nat$ and $\real$ are defined as instances of $\plus$. Moreover,
in the term $(x:\nat)+y$, the plus operator is the one defined in
the instance of $\nat:\plus$, while $x+y$ does not in general have
a definition. The term $(x:\bool)+y$ is not well typed because
$\bool$ is not defined as an instance of $\plus$.

\ifdefined\techrep
Classes in Isabelle may contain also assumptions, in addition to constants.
These are axioms for the class, and they must be proved as theorems
in instantiations. Classes can also extend existing classes. For example
we can introduce a semigroup (a set with an associative binary operation)
class as an extension of the $\plus$ class:
\[
\begin{array}{l}
\mathsf{class\ semigroup=plus\ +}\\
\qquad\mathsf{assume\ plus\_assoc:}\ x+(y+z)=(x+y)+z\\[1ex]
\mathsf{instantiation\ nat:plus}\\
\qquad\mathsf{instance\ proof\ \mbox{\ldots proof that + for \ensuremath{\nat} is associative \ldots }}
\end{array}
\]
This is a very powerful mechanism that allows for modular and reusable
developments. We can have type classes that in addition to the assumptions
contain many theorems, and when we create an instantiation of the
type class and prove the assumptions, then all these theorems become
available for the instance also.
\else
\fi

\subsection{Representation of Simulink Diagrams as Predicate Transformers}
\label{subsec:pt}

A (fragment of a) Simulink diagram is modeled intuitively  as a 
discrete symbolic transition system with input, output, current and next state. The intuition behind this representation is the following. Initially, the current state has a default value. The system representation
works in discrete steps, and, at each step, it updates the output
and the next state based on the input and the current state.

For example, an integrator block like the one from Fig.~\ref{figsimulinkaccepts} is discretized as a system parameterized by $\mathit{dt} > 0$, with input $x$ and current state $s$, and output $y:=s$ and
next state $s':=s+x\cdot\mathit{dt}$.

Formally, we model these systems in Isabelle as {\em monotonic 
predicate transformers} \cite{dijkstra-75}, mapping predicates (sets) over the output and next state into predicates (sets) over the input and current state. A monotonic predicate transformer $S$ with input $x$, current state $s$, output $y$ and next state $s'$,
for a set $q$ of pairs $(y,s')$, $S(q)$ returns the set of all pairs $(x,s)$ such that if the execution of $S$ starts
in $(x,s)$ then $S$ does not fail and results in a pair 
$(y,s')\in q$. A detailed discussion of the choice for this semantics is outside the scope of this paper, and is extensively presented in \cite{preoteasa:tripakis:2014,tripakis:2011,DBLP:conf/spin/DragomirPT16}.

In Isabelle, the integrator block is represented as the predicate transformer \rs
$$
\mathsf{Integrator}(dt)(q)(x,s) = q(s,s+x\cdot dt) \rd
$$
and it has the type $'a\to(\tp a:\plus \times \tp a \to \bool)\to 
(\tp a \times \tp a \to \bool)$. In what follows we do not make
a distinction between the input and current state, and output and
next state, respectively. In general, a Simulink diagram is modeled as a predicate transformer with input (and current state) of a type variable $\tp a$, and output (and next state)
of a type variable $\tp b$. The type of this predicate transformer is $(\tp b \to \bool) \to (\tp a \to \bool)$ and we use the notation $\tp a \ptto \tp b$ for it.
Often $\tp a$ and $\tp b$ will be Cartesian products, including the empty product ($\tunit$). We denote by $():\tunit$ the {\em empty tuple}.  

For a predicate transformer mapping, for example, inputs $(x,y,z)$ into output expressions $(x+y, x\cdot z)$, we use the notation $[x,y,z \leadsto x+y,x\cdot z]$ where \rs
$$
[x,y,z \leadsto x+y,x\cdot z](q)(x,y,z) = q(x+y,x\cdot z) \rd
$$
Using this notation, the constant and the integrator blocks become \rs
$$
\begin{array}{l}
\const(a) = [()\leadsto a], \ \ \ \ 
\mathsf{Integrator}(dt) = [x,s\leadsto s,s + x\cdot dt].
\end{array} \rd
$$ 
We denote by $\Id$ the identity predicate transformer ($[x\leadsto x]$).

A block diagram is modeled in Isabelle as an expression of predicate transformers corresponding to the basic blocks, using three composition operators: serial ($\circ$), parallel
($\para$), and feedback ($\mathsf{fb}$). The serial composition of predicate transformers is exactly the 
composition of functions. The parallel and feedback compositions are described in \cite{DBLP:conf/spin/DragomirPT16,preoteasa:tripakis:2016}. 
\ifdefined\techrep We just mention here 
that the parallel composition of $[x\leadsto e(x)]$ and
$[y\leadsto f(y)]$ is $[x,y\leadsto e(x),f(y)]$. 
\fi
For this presentation, the typing of these operations is important. The typing of the serial composition is standard. The parallel
and feedback compositions have the types: \rs
$$
\begin{array}{l}
\para : (\tp a \ptto \tp b) \to (\tp c \ptto \tp d) \to
(\tp a \times \tp c \ptto \tp b \times \tp d), \ 
\mathsf{fb}:
(\tp a \times \tp b \ptto \tp a \times \tp c) \to (\tp b \ptto \tp c)
\end{array}\rd
$$ 

Using these notations, the predicate transformer
for the rightmost diagram from Fig.~\ref{figsimulinkaccepts} is given by 
$$
  (((\const(\mathsf{s\_bool}(2)) \para \const(3)) \circ [x,y\leadsto x + y])\para \Id)\circ \mathsf{Integrator}(dt)
$$ 
We use here the $\Id$ predicate transformer to model and connect the current state of the integrator.
We also use the polymorphic function 
$\mathsf{s\_bool}$ which for $2$ returns $\True$ if the type of the result is Boolean, and
$1$ if the type of the result is real. The type of the result in this case is real
as it is inferred from the addition block (following the constant blocks).

\ifdefined\techrep
\section{Arity of Simulink Wires}
\label{sec:arity}

Simulink blocks are modeled in our Isabelle based framework by monotonic
predicate transformers. Yet, certain basic blocks require special
attention when translating them. A first example is provided in Fig. \ref{fig:mux_a}.

\begin{figure}[t]
\centering
\subfloat[Mux blocks connected in series]{
\begin{tikzpicture} 		
	\node[draw, minimum height = 5ex] (Mux) {$\mathsf{Mux_1}$};
	\node[right = 4ex of Mux.east](g){$(a,b,c,d)$};

	\node[draw, minimum height = 5ex, above left = -1ex and 10ex of Mux] (Mux1) {$\mathsf{Mux_2}$};
	\node[left = 3ex of Mux1.south west, anchor = south east](b){$b$};
	\node[left = 3ex of Mux1.north west, anchor = north east](a){$a$};

	\node[draw, minimum height = 5ex, below left = -1ex and 10ex of Mux] (Mux2) {$\mathsf{Mux_3}$};
	\node[left = 3ex of Mux2.south west, anchor = south east](d){$d$};
	\node[left = 3ex of Mux2.north west, anchor = north east](c){$c$};

	\draw[-latex'](a) -- (a -| Mux1.west);
	\draw[-latex'](b) -- (b -| Mux1.west);
	\draw[-latex'](c) -- (c -| Mux2.west);
	\draw[-latex'](d) -- (d -| Mux2.west);

	\draw[-latex'] (Mux1.east) --++ (7ex,0) node[anchor=south east] {$(a,b)$} --++ (0,-3ex) coordinate(A) -- (A -| Mux.west);
	\draw[-latex'] (Mux2.east) --++ (7ex,0) node[anchor=north east] {$(c,d)$} --++ (0,3ex) coordinate(A) -- (A -| Mux.west);

	\draw[-latex'] (Mux.east) -- (Mux.east -| g.west);
\end{tikzpicture}\label{fig:mux_a}
}
\ \ \ \ \ \ \ \ \ \ 
\subfloat[Equivalent Mux block]{
\begin{tikzpicture} 		
	\node[draw, minimum height = 13ex] (Mux) {$\mathsf{Mux}$};
	\node[left = 3ex of Mux.north west, anchor = north east](a){$a$};
	\node[below = 0.5ex of a](b){$b$};
	\node[below = 0.5ex of b](c){$c$};
	\node[below = 0.5ex of c](d){$d$};
	\node[right = 4ex of Mux.east](e) {$(a,b,c,d)$};

	\draw[-latex'](a) -- (a -| Mux.west);
	\draw[-latex'](b) -- (b -| Mux.west);
	\draw[-latex'](c) -- (c -| Mux.west);
	\draw[-latex'](d) -- (d -| Mux.west);
	\draw[-latex'] (Mux.east) -- (Mux.east -| e.west);	
\end{tikzpicture}
\label{fig:mux_b}
}
\caption{Diagrams motivating the \textit{arity} notion and computation.}
\label{fig:mux}  
\end{figure}
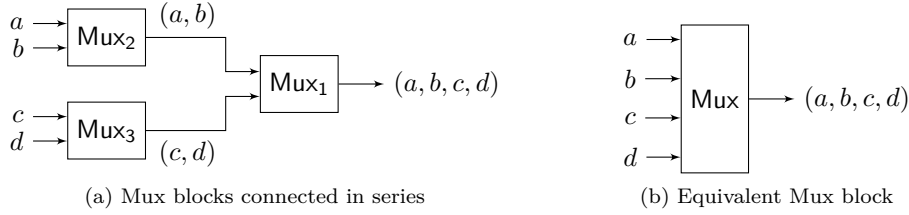

In this example, $\mathsf{Mux_2}$ groups together wires $a$ and $b$,
and outputs them as a tuple. Similarly, $\mathsf{Mux_3}$ outputs the
tuple $(c,d)$. Then $\mathsf{Mux_1}$ flattens the tuple of inputs
into one output tuple $(a,b,c,d)$. The same behavior is modeled in
Fig. \ref{fig:mux_b}, where the 3 blocks $\mathsf{Mux_1}$, $\mathsf{Mux_2}$,
and $\mathsf{Mux_3}$ are collapsed into just one $\mathsf{Mux}$.

A possible translation of the diagrams from Fig.~\ref{fig:mux} into
Isabelle is: \rs
\[
\begin{array}{lll}
\mathsf{(Mux_2\para Mux_3)\circ Mux_1} & \mbox{ and } & \mathsf{Mux}\end{array} \rd
\]
where $\mathsf{Mux_i}=[x,y\leadsto x,y]$ and $\mathsf{Mux}=[a,b,c,d\leadsto a,b,c,d]$.
Due to the parallel composition definition, we have $\mathsf{(Mux_2\para Mux_3)\circ Mux_1}=[(a,b),(c,d)\leadsto(a,b),(c,d)]$,
which is different from $\mathsf{Mux}$. 

In order to solve this  
problem, we consider that all (sub)diagrams have as input and ouput tuples
of single values, and not tuples of tuples. The parallel composition
is one construct which does not preserve this property. Ideally we
would define the parallel composition such that it splits the input
tuple in two and concatenates the output tuples into a single one.
Unfortunately, this cannot be easily achieved in Isabelle, because
concatenation of tuples of arbitrary length cannot be expressed. 

The solution is to use the parallel composition together with explicit
splitting of the input and concatenation of the output: $[(a,b,c,d)\leadsto(a,b),(c,d)]\circ(\mathsf{Mux_2\para Mux_3})\circ[(a,b),(c,d)\leadsto(a,b,c,d)]$,
and use $\Id$ for all mux blocks. Using this technique the translation
of the first diagram is \rs
\[
[(a,b,c,d)\leadsto(a,b),(c,d)]\circ(\Id\para\Id)\circ[(a,b),(c,d)\leadsto(a,b,c,d)]\circ\Id=\Id \rd
\]
and this is now equal to $\mathsf{Mux}$. However, to be able to do
this we need to know how many signals are transferred on wires. This
is not a local property (it does not depend only on the type of the
current block). For example if we analyze the diagram from Fig.~\ref{fig:mux_a},
we see that $\mathsf{Mux_1}$ has as output a tuple with four elements,
and this can be inferred only by analyzing the entire diagram. So
for every wire we need to compute its \emph{arity} (the number of
transferred signals). 

The algorithm for computing the arities is given in Fig.~\ref{fig:algo}
using classes to represent blocks in a Python style syntax. A diagram
is represented as an instance $x$ of the $\mathsf{subsystem}$ class
that contains a list of basic blocks or subsystems ($x.\mathsf{blocks}$).
Each block $x$ (basic or subsystem) has a list of input ports and
a list of output ports ($x.\mathsf{inport}$ and $x.\mathsf{outport}$).
All ports $p$ have arities ($p.\mathsf{arity}$), and initially they
are all set to $1$. A wire between two blocks $x,y$ in the diagram
is represented by a port which is output port for $x$ and input port
for $y$. The idea of the algorithm is that it updates for every block 
the arities of the output ports, based on the arities of the input
ports (function $x.\mathsf{update\_arity()}$). The function returns
$\True$ if a change occurred, and $\False$ otherwise. Because of
how wires are implemented, a change in the arity of an output port
would result in a change in the arity of an input port of a connected
block. We compute the arities of the diagram $x$ using function $x.\mathsf{compute\_arity()}$.
We iterate $x.\mathsf{update\_arity}()$ until the\emph{ computation
converges} (there are no more changes) or until we exceed a bound
on the number of iterations. The function $x.\mathsf{compute\_arity()}$
returns $\True$ if the computation converges and $\False$ otherwise.
The bound must be chosen such that the computation converges if and
only if it converges within this bound. Using the total number of
basic blocks in the model as the bound guarantees this property,
regardless of the ordering of blocks.

\ifdefined\techrep 
\begin{figure}
\centering
\begin{minipage}[t]{0.50\columnwidth}%
\textsf{\textbf{class}~block:}{\par}

\textsf{~~~~\textbf{def}~update\_arity(self):}{\par}

\textsf{~~~~~~~~\textbf{return}~True}{\par}

\textsf{~}{\par}

\textsf{~~~~\textbf{def}~compute\_arity(self, bound):}{\par}

\textsf{~~~~~~~~i = 0}{\par}

\textsf{~~~~~~~~up = True}{\par}

\textsf{~~~~~~~~\textbf{while}~up~\textbf{\and}~i < bound:}{\par}

\textsf{~~~~~~~~~~~~up = self.update\_arity()}{\par}

\textsf{~~~~~~~~~~~~i =i+1 }{\par}

\textsf{~~~~~~~~\textbf{return}~i < bound}{\par}%
\end{minipage}%
\begin{minipage}[t]{0.44\columnwidth}%
\textsf{\textbf{class}~mux(block):}{\par}

\textsf{~~~~\textbf{def}~update\_arity(self):}{\par}

\textsf{~~~~~~~~a = 0}{\par}

\textsf{~~~~~~~~\textbf{for}~x~\textbf{in}~inport: a = a + x.arity}{\par}

\textsf{~~~~~~~~up = (a != outport{[}0{]}.arity)}{\par}

\textsf{~~~~~~~~outport{[}0{]}.arity = a }{\par}

\textsf{~~~~~~~~\textbf{retun} up}{\par}

\textsf{~}{\par}

\textsf{\textbf{class} subsystem(block):}{\par}

\textsf{~~~~\textbf{def} update\_arity(self):}{\par}

\textsf{~~~~~~~~up = False}{\par}

\textsf{~~~~~~~~\textbf{for} b \textbf{in} self.blocks:}{\par}

\textsf{~~~~~~~~~~~~up = up \textbf{or} b.update\_arity()}{\par}

\textsf{~~~~~~~~\textbf{return} up}{\par}%
\end{minipage}
\caption{Algorithm for computing wire arities}
\label{fig:algo} 
\end{figure}
\else 
\begin{figure}
\centering
\begin{minipage}[t]{0.50\columnwidth}%
\textsf{\textbf{\footnotesize{}class}}\textsf{\footnotesize{}~block:}{\footnotesize \par}

\textsf{\footnotesize{}~~~~}\textsf{\textbf{\footnotesize{}def}}\textsf{\footnotesize{}
update\_arity(self):}{\footnotesize \par}

\textsf{\footnotesize{}~~~~~~~~}\textsf{\textbf{\footnotesize{}return}}\textsf{\footnotesize{}
True}{\footnotesize \par}

\textsf{\footnotesize{}~}{\footnotesize \par}

\textsf{\footnotesize{}~~~~}\textsf{\textbf{\footnotesize{}def}}\textsf{\footnotesize{}
compute\_arity(self, bound):}{\footnotesize \par}

\textsf{\footnotesize{}~~~~~~~~i = 0}{\footnotesize \par}

\textsf{\footnotesize{}~~~~~~~~up = True}{\footnotesize \par}

\textsf{\footnotesize{}~~~~~~~~}\textsf{\textbf{\footnotesize{}while}}\textsf{\footnotesize{}
up }\textsf{\textbf{\footnotesize{}and}}\textsf{\footnotesize{}
i < bound:}{\footnotesize \par}

\textsf{\footnotesize{}~~~~~~~~~~~~up = self.update\_arity()}{\footnotesize \par}

\textsf{\footnotesize{}~~~~~~~~~~~~i =i+1 }{\footnotesize \par}

\textsf{\footnotesize{}~~~~~~~~}\textsf{\textbf{\footnotesize{}return}}\textsf{\footnotesize{}
i < bound}{\footnotesize \par}%
\end{minipage}%
\begin{minipage}[t]{0.44\columnwidth}%
\textsf{\textbf{\footnotesize{}class}}\textsf{\footnotesize{} mux(block):}{\footnotesize \par}

\textsf{\footnotesize{}~~~~}\textsf{\textbf{\footnotesize{}def}}\textsf{\footnotesize{}
update\_arity(self):}{\footnotesize \par}

\textsf{\footnotesize{}~~~~~~~~a = 0}{\footnotesize \par}

\textsf{\footnotesize{}~~~~~~~~}\textsf{\textbf{\footnotesize{}for}}\textsf{\footnotesize{}
x }\textsf{\textbf{\footnotesize{}in}}\textsf{\footnotesize{} inport:
a = a + x.arity}{\footnotesize \par}

\textsf{\footnotesize{}~~~~~~~~up = (a != outport{[}0{]}.arity)}{\footnotesize \par}

\textsf{\footnotesize{}~~~~~~~~outport{[}0{]}.arity = a }{\footnotesize \par}

\textsf{\footnotesize{}~~~~~~~~}\textsf{\textbf{\footnotesize{}retun}}\textsf{\footnotesize{}
up}{\footnotesize \par}

\textsf{\footnotesize{}~}{\footnotesize \par}

\textsf{\textbf{\footnotesize{}class}}\textsf{\footnotesize{} subsystem(block):}{\footnotesize \par}

\textsf{\footnotesize{}~~~~}\textsf{\textbf{\footnotesize{}def}}\textsf{\footnotesize{}
update\_arity(self):}{\footnotesize \par}

\textsf{\footnotesize{}~~~~~~~~up = False}{\footnotesize \par}

\textsf{\footnotesize{}~~~~~~~~}\textsf{\textbf{\footnotesize{}for}}\textsf{\footnotesize{}
b }\textsf{\textbf{\footnotesize{}in}}\textsf{\footnotesize{} self.blocks:}{\footnotesize \par}

\textsf{\footnotesize{}~~~~~~~~~~~~up = up }\textsf{\textbf{\footnotesize{}or}}\textsf{\footnotesize{}
b.update\_arity()}{\footnotesize \par}

\textsf{\footnotesize{}~~~~~~~~}\textsf{\textbf{\footnotesize{}return}}\textsf{\footnotesize{}
up}{\footnotesize \par}%
\end{minipage}
\caption{Algorithm for computing wire arities}
\label{fig:algo} 
\end{figure}
\fi

However, there are diagrams for which the computation does not converge. An
example is illustrated in Fig. \ref{fig:inf_loop}. In this diagram,
the arities of the input and output ports change continuously due
to the feedback wire from $\mathsf{Delay}$ to $\mathsf{Mux}$. This
diagram is also not accepted by Simulink for simulation for the same reason.

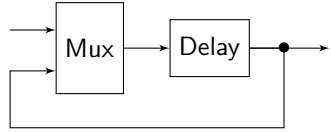
\begin{figure}
\centering
\begin{tikzpicture} 		
	\node[minimum height = 5ex] (Cst) {};
	\node[draw, minimum height = 8ex, below right=-5ex and 4ex of Cst] (Mux) {$\mathsf{Mux}$};
	\node[draw, minimum height = 5ex, right=4ex of Mux] (Delay) {$\mathsf{Delay}$};

	\draw[-latex'] (Cst.east) --++ (2ex,0) coordinate(A) --(A -| Mux.west);
	\draw[-latex'] (Mux) -- (Delay);
	\draw[-latex'](Delay.east)-- ++(3ex,0) node(C){$\bullet$} -- ++(0ex,-7ex) coordinate(A) 
		-- (A -| Mux.west) -- ++ (-4ex, 0) -- ++(0, 5ex) coordinate(B) -- (B -| Mux.west);
	\draw[-latex'](Delay.east) -- ++(7ex, 0);
\end{tikzpicture}

\caption{An infinite-loop diagram.}
\label{fig:inf_loop}
\end{figure}

An alternative solution to the translation of mux blocks would be to use lists of values instead of tuples. Then we can define
more easily the parallel composition (concatenation of lists is associative), but we would
loose the typing information for individual inputs and outputs. In this situation we would need
to use a single type that contains all values possible in a diagram (Booleans, reals).
Another drawback of this approach would be also that we could not detect statically
problems that occur when the algorithm from Fig.~\ref{fig:algo} does not  
converge (like in Fig.~\ref{fig:inf_loop}).
By using a system like PVS that supports dependent types, we could improve 
the list representation of inputs and outputs to account for specific types of individual
inputs. Yet, the convergence problem of the arity calculation needs to be solved by
proving a type constraint condition theorem in these systems.
\fi

\section{Constant Blocks}
\label{sec:constants}

Simulink diagrams may contain constant blocks, parameterized by numeric
constants. These are blocks without input and with one single output
which is always equal to the constant's parameter. By default, Simulink
constants do not have associated types. In order to have the possibility
to instantiate these types later for reals, integers, Booleans, or
other types, we use uninterpreted constants. By default, numeric constants
in Isabelle are polymorphic. If no type is explicitly set to a constant
in a term $t=12$, then Isabelle associates a type variable $\tp a:\num$
to this constant. If the term is used in a context where the type
is more specific ($t=12\land\suc(t)=t'$) then Isabelle uses the type
class instantiation to the specific type (in this case natural because
of the successor function).

\begin{figure}[t]
\centering
\subfloat[]{\begin{tikzpicture}
\node[draw,minimum width = 5ex, minimum height = 4ex](a){$1$};
\node[draw,minimum width = 5ex, minimum height = 4ex,below = 1ex of a](b){$2$};
\node[fit = (a) (b)](c){};
\node[draw, minimum height = 8ex, right = 3ex of c](d){$\not=$};
\draw[-latex'](a) -- (a -| d.west);
\draw[-latex'](b) -- (b -| d.west);
\draw[-latex'](d.east) -- ++ (3ex, 0);
\node[draw, inner sep = 1ex, fit= (a) (b) (d)](e){};
\node[below = 1ex of e]{$\mathsf{Compare}$};
\end{tikzpicture}
\label{fig:const-order}
}
\ \ \ \ \ 
\subfloat[]{
\begin{tikzpicture}
\node[draw,minimum width = 4ex, minimum height = 4ex](a){$1$};
\node[draw,minimum width = 4ex, minimum height = 4ex,below = 1ex of a](b){$2$};
\node[fit = (a) (b)](c){};
\node[draw, minimum height = 8ex, right = 3ex of c](d){$\not=$};
\draw[-latex'](a) -- (a -| d.west);
\draw[-latex'](b) -- (b -| d.west);
\node[draw, inner sep = 1ex, fit= (a) (b) (d)](e){};
\node[below = 1ex of e]{$\mathsf{Compare}$};
\node[draw, inner ysep = 2ex, right = 9ex of e](f){$\And$};
\draw[-latex'](d.east)++(0,1ex)coordinate(A) -- (A-|f.west);
\node[draw, right = 2ex of e,
  minimum width = 4ex, minimum height = 4ex, anchor = north west](g){1};
\draw[-latex'](g.east)coordinate(A) -- (A-|f.west);
\draw[-latex'](f.east) -- ++(3ex,0);
\end{tikzpicture}
\label{fig:const-order-and}
}
\ \ \ 
\raisebox{4pt}{\subfloat[]{
\begin{tikzpicture}
\node[draw,minimum width = 5ex, minimum height = 4ex](a){$1.5:\real$};
\node[draw,minimum width = 5ex, minimum height = 4ex,below = 1ex of a](b){$1:\bool$};
\node[draw,minimum width = 5ex, minimum height = 4ex,below = 1ex of b](c){$3$};
\node[fit = (a) (b) (c)](d){};
\node[draw, minimum height = 8ex, right = 4ex of d](e){$\And$};
\draw[-latex'](b.east) --++(2ex,0)node[anchor = south]{$y$} -- (b -| e.west);
\draw[-latex'](a.east) -- ++(3ex,0)node[anchor = west]{$x$} -- ++(0,-2.5ex)coordinate(A) -- (A-| e.west);
\draw[-latex'](c.east) -- ++(3ex,0)node[anchor = west]{$z$} -- ++(0,2.5ex)coordinate(A) -- (A-| e.west);
\draw[-latex'](e.east) -- ++(3ex,0);
\end{tikzpicture}
\label{fig:conj}
}}
\caption{(a) Comparison on constants, (b) Comparison into conjunction,
(c) $\And$ on typed constants} \ifdefined\techrep \else \vspace{-3ex}
\fi 
\end{figure}
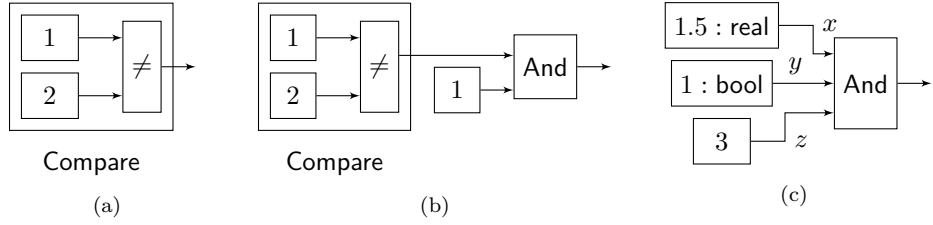

Due to this polymorphic treatment of constants, in some contexts it arises the problem that 
the types of these constants are not part of the type of the resulting predicate 
transformer. Consider for example
the diagram from Fig.~\ref{fig:const-order}. The Isabelle definition for this diagram is \rs
\[
\mathsf{Compare}=(\const(1:\tp a:\num)\para\const(2))\circ[x,y\leadsto x\not=y]=[()\leadsto 1\not=2] \rd
\]

In this definition $\tp a$ is the inferred type of constants $1$ and
$2$. The problem with this definition is that the type $\tp a$ is not
part of the type of $\mathsf{Compare:\tunit\ptto\bool}$. If this
definition would be allowed, then we will have an unsound system,
because for example if $\tp a$ is instantiated by $\real$, then $(1:\real)\not=2$
is true and $\mathsf{Compare}=[()\leadsto\True]$, but if $\tp a$ is
instantiated by $\tunit$, then $(1:\tunit)\not=2$ is false ($\tunit$
contains only one element) and $\mathsf{Compare}=[()\leadsto\False]$,
and we can derive $[()\leadsto\False]=[()\leadsto\True]$ which is
false. In order to instantiate $\tunit$ for $\tp a$, the type $\tunit$
must be of class $\mathsf{numeral}$. Although by default this is not the case in Isabelle,
we can easily add an instantiation of $\tunit$ as  $\mathsf{numeral}$ and obtain
this contradiction.

Isabelle allows this kind of definition, but it gives a warning message
(``Additional type variable(s) in specification of $\mathsf{Compare}:\tp a:\num$''),
and it defines the function $\mathsf{Compare}$ to depend on an additional
type variable: \rs
\[
\mathsf{Compare}(\tp a:\num)=(\const(2:\tp a)\para\const(1))\circ[x,y\leadsto x\not=y] \rd
\]
Now $\mathsf{Compare}(\real)$ and $\mathsf{Compare}(\tunit)$ are
different terms, so they are not equal anymore and we cannot derive
$[()\leadsto\False]=[()\leadsto\True]$. Assume now that we compose
the $\mathsf{Compare}$ block with a conjunction block as in Fig.~\ref{fig:const-order-and}. \rs
\[
A=(\mathsf{Compare}\parallel\const(1))\circ\And \rd
\]
However, this definition is now incorrect because $\mathsf{Compare}$
has an additional type parameter. The correct definition would be: \rs
\[
A(\tp a:\num)=(\mathsf{Compare}(\tp a)\parallel\const(1))\circ\And \rd
\]
When we generate the definition for the diagram from Fig.~\ref{fig:const-order-and}
we do not know that $\mathsf{Compare}$ needs the additional type
parameter. To have control over the type parameters we add them systematically
for all constants occurring in the diagram. 
\ifdefined\techrep \else
Moreover, we define the constants with a variable parameter. Due to the lack of space,
the rationale for these definitions is discussed in \cite{preoteasa:dragomir:tripakis:2016}.
\fi

\hide{
\ifdefined\techrep
\begin{figure}[t]
\centering
\subfloat[]{
\begin{tikzpicture}
\node[draw, minimum width = 4ex, minimum height = 4ex](a){$1$};
\node[draw, minimum width = 4ex, minimum height = 4ex, below = 1ex of a](b){$2$};
\node[draw, minimum width = 4ex, minimum height = 4ex, below = 1ex of b](c){$3$};
\node[fit = (a) (b)](d){};
\node[draw, minimum width = 4ex, minimum height = 8ex, right = 3ex of d](add){$\Add$};
\draw[-latex'](a.east) -- (a -| add.west);
\draw[-latex'](b.east) -- (b -| add.west);
\draw[-latex'](add.east) -- ++(3ex,0) coordinate(A);
\draw[-latex'](c.east) -- (c -| A);
\end{tikzpicture}
\label{fig:const-add}
}
\ \ \ \ \ \ \ \ \ \ 
\subfloat[]{
\begin{tikzpicture}
\node[draw,minimum width = 5ex, minimum height = 4ex](a){$1.5:\real$};
\node[draw,minimum width = 5ex, minimum height = 4ex,below = 1ex of a](b){$1:\bool$};
\node[draw,minimum width = 5ex, minimum height = 4ex,below = 1ex of b](c){$3$};
\node[fit = (a) (b) (c)](d){};
\node[draw, minimum height = 8ex, right = 4ex of d](e){$\And$};
\draw[-latex'](b.east) --++(2ex,0)node[anchor = south]{$y$} -- (b -| e.west);
\draw[-latex'](a.east) -- ++(3ex,0)node[anchor = west]{$x$} -- ++(0,-2.5ex)coordinate(A) -- (A-| e.west);
\draw[-latex'](c.east) -- ++(3ex,0)node[anchor = west]{$z$} -- ++(0,2.5ex)coordinate(A) -- (A-| e.west);
\draw[-latex'](e.east) -- ++(3ex,0);
\end{tikzpicture}
\label{fig:conj}
}
\caption{(a) Addition of Constants, (b) $\mathsf{And}$ on typed Constants.} 
\end{figure}
\fi
}

\ifdefined\techrep
However, using just type variables
as parameters still results in several problems explained next. Assume again that
we have three constants as in Fig.~\ref{fig:const-order-and}. 
We define the three constant blocks as follows: \rs
\[
\begin{array}{lll}
\mathsf{ConstA}(\tp a) & = & \const(1:\tp a) \\
\mathsf{ConstB}(\tp b) & = & \const(2:\tp b) \\
\mathsf{ConstC}(\tp c) & = & \const(1:\tp c) 
\end{array} \rd
\]
and the diagram from Fig.~\ref{fig:const-order-and} as: \rs
\[
A(\tp a,\tp b,\tp c) = (((\mathsf{ConstA}(\tp a)\para\mathsf{ConstB}(\tp b))\circ
[x,y\leadsto x\not = y])\parallel\mathsf{ConstC}(\tp c))\circ \And \rd
\]
The problem with this definition is that both outputs of $\mathsf{ConstA}$
and $\mathsf{ConstB}$ are entering the same comparison block. This
implies that the two constants must be of the same type. However,
since we used the explicit type variables $\tp a$ and $\tp b$, Isabelle
 cannot unify them. 
An alternative definition is to use the same type for
all constants: \rs
\[
A(\tp a) = (((\mathsf{ConstA}(\tp a)\para\mathsf{ConstB}(\tp a))\circ
[x,y\leadsto x\not = y])\parallel\mathsf{ConstC}(\tp a))\circ \And \rd
\]
We would need to use the same type for all three constants because we do not know during translation that
only the types of $\mathsf{ConstA}$ and $\mathsf{ConstB}$ should be unified. This information could be extracted 
through type inference during translation, however our aim is to use Isabelle's mechanism for this feature.
The drawback of this solution is that although, in the original diagram,
the type of constant $\mathsf{ConstC}$ is independent of the types
of $\mathsf{ConstA}$ and $\mathsf{ConstB}$, in this definition they
are unified. 
We would like to obtain a definition of
$A$ where the types of constants $\mathsf{ConstA}$ and $\mathsf{ConstB}$
are unified, and the type of $\mathsf{ConstC}$ is independent. For
this we define the constants with a variable parameter \rs
\else
With this method the constant blocks from Fig.~\ref{fig:const-order-and} are defined by \rs
\fi
\begin{equation}
\ifdefined\techrep
\begin{array}{lll}
\mathsf{ConstA}(x:\tp a) & = & \const(1:\tp a) \\
\mathsf{ConstB}(y:\tp b) & = & \const(2:\tp b) \\
\mathsf{ConstC}(z:\tp c) & = & \const(1:\tp c) 
\end{array} \rd
\else
\begin{array}{lll}
\mathsf{ConstA}(x:\tp a)  =  \const(1:\tp a) & \mbox{ and } &
\mathsf{ConstB}(y:\tp b)  =  \const(2:\tp b) \mbox{ and }\\
\mathsf{ConstC}(z:\tp c)  =  \const(1:\tp c)
\end{array} \rd
\fi
\label{eq:const-def}
\end{equation}
and the diagram is defined by \rs
\begin{equation}
A(x,y,z) = (((\mathsf{ConstA}(x)\para\mathsf{ConstB}(y))\circ[x,y\leadsto x\not=y])\parallel\mathsf{ConstC}(z))\circ\And \label{eq:dgr-vars}  \rd
\end{equation}
In this approach, variables $x,y,z$ are used only to control the types of the
constants. In this definition, because
outputs of $\mathsf{ConstA}$ and $\mathsf{ConstB}$ are entering
the comparison block, the types of $x$ and $y$ are unified. If we need an instance
of $A$ for type real for constants $\mathsf{ConstA}$ and $\mathsf{ConstB}$
and type Boolean for $\mathsf{ConstC}$, then we can specify it using
the term $A(x:\real,y:\real,z:\bool)$.

This definition mechanism is implemented in our Simulink to Isabelle model translator
under the $\mathsf{-const}$ option. When the option is set, then the constants are
defined as in (\ref{eq:const-def}), and the diagrams using these
constants are defined as in (\ref{eq:dgr-vars}). When the option
is not given, then the constants are defined as in: \rs
\[
\begin{array}{lllll}
\mathsf{ConstA} = \const(1) & \mbox{ and } &
\mathsf{ConstB} = \const(2) & \mbox{ and } &
\mathsf{ConstC} = \const(1)
\end{array} \rd
\]
and they are used as in: $A = (((\mathsf{ConstA}\para\mathsf{ConstB})\circ[x,y\leadsto x\not=y])\parallel\mathsf{ConstC})\circ\And$.
When the constant blocks in a Simulink diagram define an output type, we
simply use them as in $\const(1.5:\real)$ (Fig.~\ref{fig:conj}).

\section{Conversion Blocks}
\label{sec:conversion}

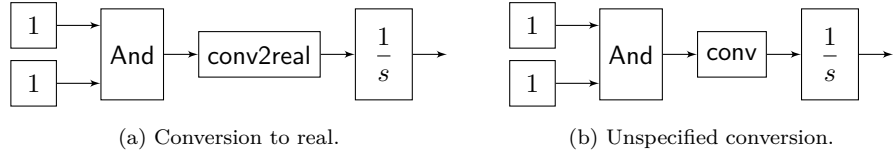
\begin{figure}[t]
\centering
\subfloat[Conversion to real.]{\begin{tikzpicture}
\node[draw, minimum width = 4ex, minimum height = 4ex](a){$1$};
\node[draw, minimum width = 4ex, minimum height = 4ex, below = 1ex of a](b){$1$};
\node[fit = (a) (b)](d){};
\node[draw, minimum width = 4ex, minimum height = 8ex, right = 3ex of d](aa){$\And$};
\draw[-latex'](a.east) -- (a -| aa.west);
\draw[-latex'](b.east) -- (b -| aa.west);

\node[draw,minimum width = 5ex, minimum height = 4ex, right = 3ex of aa](conv){$\mathsf{conv2real}$};
\node[draw,minimum width = 5ex, minimum height = 8ex, right = 3ex of conv](inte){$\displaystyle\frac{1}{s}$};

\draw[-latex'](aa) -- (conv);
\draw[-latex'](conv) -- (inte);
\draw[-latex'](inte.east) -- ++(3ex,0);
\end{tikzpicture}
\label{fig:conv-to-real}
}
\ \ \ \ \ 
\subfloat[Unspecified conversion.]{\begin{tikzpicture}
\node[draw, minimum width = 4ex, minimum height = 4ex](a){$1$};
\node[draw, minimum width = 4ex, minimum height = 4ex, below = 1ex of a](b){$1$};
\node[fit = (a) (b)](d){};
\node[draw, minimum width = 4ex, minimum height = 8ex, right = 3ex of d](aa){$\And$};
\draw[-latex'](a.east) -- (a -| aa.west);
\draw[-latex'](b.east) -- (b -| aa.west);

\node[draw,minimum width = 5ex, minimum height = 4ex, right = 3ex of aa](conv){$\mathsf{conv}$};
\node[draw,minimum width = 5ex, minimum height = 8ex, right = 3ex of conv](inte){$\displaystyle\frac{1}{s}$};

\draw[-latex'](aa) -- (conv);
\draw[-latex'](conv) -- (inte);
\draw[-latex'](inte.east) -- ++(3ex,0);
\end{tikzpicture}
\label{fig:conversion}
}
\caption{Conversions examples.}
\label{fig:conv} \ifdefined\techrep \else \vspace{-4ex} 
\fi
\end{figure}

Simulink diagrams may also contain conversion blocks. The type of
the input of a conversion is inherited and the type of the output
is usually specified (Boolean, real, \ldots). However we can have also
situations when the output is not specified, and it is inherited from
the type of the inputs of the block that follows a conversion. In Fig.~\ref{fig:conv-to-real} we
illustrate an explicit conversion to real, while Fig.~\ref{fig:conversion} presents an unspecified conversion. 

As with the other blocks we want to define these conversions locally,
without knowing the types of the inputs and outputs,
when the output type is unspecified. In doing so, we use the overloading
mechanism of Isabelle. Overloading is a feature that allows using
the same constant name with different types. For the conversion blocks we introduce
the following definitions. \rs
\ifdefined\techrep
\[
\begin{array}{l}
\mathsf{consts\ conv}:\tp a\to\tp b\\
\mathsf{overloading}\\
\qquad\mathsf{conv\_id}=(\mathsf{conv}:\tp a\to\tp a)\\
\qquad\mathsf{conv\_bool\_real}=(\mathsf{conv}:\bool\to\real)\\
\qquad\mathsf{conv\_real\_bool}=(\mathsf{conv}:\real\to\bool)\\
\mathsf{begin}\\
\qquad\mathsf{conv\_id}(x:\tp a):=x\\
\qquad\mathsf{conv\_bool\_real}(x:\bool):=\mathsf{if}\ x\ \mathsf{then}\ 1\ \mathsf{else}\ 0\\
\qquad\mathsf{conv\_real\_bool}(x:\real):=(x\not=0)\\
\mathsf{end}
\end{array} \rd
\]
\else
\[
\begin{array}{l}
\mathsf{consts\ conv}:\tp a\to\tp b\\
\mathsf{overloading}\\
\qquad\mathsf{conv}(x:\tp a):=x\\
\qquad\mathsf{conv}(x:\bool):=\mathsf{if}\ x\ \mathsf{then}\ (1:\real)\ \mathsf{else}\ 0\\
\qquad\mathsf{conv}(x:\real):=(x\not=0) 
\end{array}\rd
\]
\fi
This definition introduces an arbitrary function $\mathsf{conv}$
from a type variable $\tp a$ to a type variable $\tp b$, and it also defines
three overloadings for this function. The term $\mathsf{conv}(x)$
in general is of type $\tp b$ and $x$ is of type $\tp a$. If we restrict
the type $\tp a$ and $\tp b$ to $\real$ and $\bool$, then we have \rs
\ifdefined\techrep
\[
(\mathsf{conv}(x:\real):\bool)=\mathsf{conv\_real\_bool}(x)=(x\not=0) \rd
\]
\else
\[
(\mathsf{conv}(x:\real):\bool)=(x\not=0) \rd
\]
\fi
When we translate a conversion block, if we know the output type,
then we use the conversion restricted to this output type, otherwise
we use the unrestricted conversion. For example the conversion from
\ref{fig:conv-to-real} is translated into $[x\leadsto(\mathsf{conv}(x):\real)]$.
The entire diagram from Fig.~\ref{fig:conv-to-real} is translated into \rs
\begin{equation}
(((\const(1)\parallel\const(1))\circ\And\circ[x\leadsto(\mathsf{conv}(x):\real)])\para\Id)\circ[x,s\leadsto s,s+x\cdot\mathit{dt}]\label{eq:conv-inte} \rd
\end{equation}
The identity block ($\Id$) is used here for the current state input of the
integral block.
The conversion from Fig.~\ref{fig:conversion} is translated into
$[x\leadsto\mathsf{conv}(x)]$. This diagram becomes  \rs
\[
(((\const(1)\parallel\const(1))\circ\And\circ[x\leadsto\mathsf{conv}(x)])\para\Id)\circ[(x:\real),s\leadsto s,s+x\cdot\mathit{dt}] \rd
\]
and compared with (\ref{eq:conv-inte}) the only difference is that
in the later case, the type of the output of the conversion is not
specified. However, in both cases, the inputs of the conversions must
be Boolean because of the $\And$ block, and the outputs must be $\real$
because of the integral block. In both cases the translations are
equivalent to $[(s:\real)\leadsto s,s+\mathit{dt}]$.

\section{Boolean Blocks}
\label{sec:boolean}

Simulink Boolean blocks are also challenging to implement due to the
fact that, for example, the inputs to a conjunction block could have
different types (real, Boolean, unspecified), as illustrated in Fig.~\ref{fig:conj}. 
In languages that allow it (e.g., C, Python), it is common practice to use 
numerical values in Boolean expressions, with the meaning that
non-zero is true. Similarly, it is common practice to use Boolean values in 
numeric expressions. Simulink also allows these cases, but Isabelle does not. We show in this and next section how to solve these problems.

Consider the example from Fig.~\ref{fig:conj}. If we would simply take the conjunction of all 
inputs \rs
\[
(\const(1.5:\real)\para\const(1:\bool)\para\const(3))\circ[x,y,z\leadsto x\land y\land z] \rd
\]
we will obtain in Isabelle a type error, because $x$ has type $\real$, $y$ has type
$\bool$ and $z$ has type $\tp a:\num$, and their conjunction is not
well typed.

To fix this typing problem, we implement the conjunction block in the following
way: $\And = [x,y,z\leadsto(x\not=0)\land(y\not=0)\land(z\not=0)]$.
In this expression the types of variables $x$, $y$, and $z$ are
independent of each other, and also of the Boolean output, and they
can match the types of the blocks that are input to $\And$. There
are still some details to consider. If input $x$ is real,
then $x\not=0$ is true if and only if $x$ is not zero, and this
coincides with the semantics of $\And$ in Simulink. However, if 
the input $y$ is Boolean, then the expression $y\not=0$
is not well typed, unless we add additional class instantiation
in Isabelle: \rs
\[
\begin{array}{l}
\mathsf{instantiation\ bool:zero=}\\
\qquad(0:\bool):=\False
\end{array} \rd
\]
Intuitively this instantiation provides the interpretation of constant
$0$ as $\False$, when $0$ is used as a Boolean value. With this
the expression $(y:\bool)\not=0$ is equivalent to $y\not=\False$
and it is equivalent to $y$. 
The same holds for the expression $1:\mathsf{bool}$ which is not well 
typed unless we provide an instantiation of $\mathsf{bool}$ as $\mathsf{numeral}$,
where every (non-zero) numeral constant is $\True$.
These definitions formalize the behavior described by Simulink in its documentation.

Using this approach, the translation of the diagram from Fig.~\ref{fig:conj}
is: \rs
\[
(\const(1.5:\real)\para\const(1:\bool)\para\const(3))\circ[x,y,z\leadsto x\not=0\land y\not=0\land z\not=0] \rd
\]
and it is equal to \rs
\[
(\const(1.5:\real)\para\const(\True)\para\const(3))\circ[x,y,z\leadsto x\not=0\land y\land z\not=0]
\]
because $y$ is of type $\mathsf{bool}$ and $(y\not=0) = y$.  If we expand the serial composition and simplify the term, we obtain 
$[()\leadsto(3:\tp a:\{\num,\zero\})\not=0]$.
The equality $(3:\tp a:\{\num,\zero\})\not=0$ cannot be simplified.
This is because the type $\tp a:\{\num,\zero\}$ has all numeric constants
$1,2,\ldots$ ($\num$) and the constant $0$ ($\zero$), but no relationship
between these constants is known. If we know that we only use the
type $\tp a$ with instances where the numeric constants $1,2,\ldots$
are always different from $0$, then we can create a new class based
on $\num$ and $\zero$ that has also the property that $n\not=0$
for all $n\in\{1,2,\ldots\}$. Formally we can introduce this class
in Isabelle by \rs
\[
\begin{array}{l}
\mathsf{class\ numeral\_nzero=zero+numeral\ +}\\
\qquad\mathsf{assume\ numeral\_nzero[simp]:\ }(\forall a.\mathsf{numeral}(a)\not=0)
\end{array} \rd
\]
The new class $\mathsf{numeral\_nzero}$ contains the numeric constants
$\{0,1,2,\ldots\}$ but also it has the property that all numbers
$1,2,\ldots$ are different from $0$ $(\forall a.\mathsf{numeral}(a)\not=0)$.
In this property $a$ ranges over the binary representations of the
numbers $1,2,\ldots$. This property is called $\mathsf{numeral\_nzero}$,
and the $\mathsf{[simp]}$ declaration tells Isabelle to use it
automatically as simplification rule. Now the equality $(3:\tp a:\mathsf{numeral\_nzero})\not=0$
is also automatically simplified to $\True$. 

We provide the following class instantiation: \rs
\[
\begin{array}{l}
\mathsf{instantiation\ \bool:numeral\_nzero}=\\
\qquad(0:\bool):=\False\ \ | \ \ 
(\num(a):\bool):=\True
\end{array} \rd
\]
Because in this class we have also the assumption $(\forall a.\mathsf{numeral}(a)\not=0)$,
we need to prove it, and it trivially holds because $\False\not=\True$.
Similarly we need to introduce instantiations of $\mathsf{numeral\_nzero}$
to real, integer, and natural numbers. In these cases, since real,
integer, and natural are already instances of $\num$ and $\zero$,
we do not need to define $0$ and $\num(a)$, but we only need to prove
the property $(\forall a.\mathsf{numeral}(a)\not=0)$. 

With this new
class, the translation of diagram from Fig.~\ref{fig:conj} becomes: \rs
\[
\begin{array}{c}
(\const(1.5:\real)\para\const(1:\bool)\para\const(3:\tp a:\mathsf{numeral\_nzero})) \ \circ \\
~[x,y,z\leadsto x\not=0 \land y\not=0 \land z\not=0]
\end{array} \rd
\]
Because of the properties of types $\real$, $\bool$, and $\tp a:\mathsf{numeral\_nzero}$,
it is equal to 
\[
\begin{array}{c}
(\const(1.5:\real)\para\const(\True)\para\const(3:\tp a:\mathsf{numeral\_nzero}))\ \circ \\
~[x,y,z\leadsto x\not=0 \land y  \land z\not=0]
\end{array} \rd
\]
and, after expanding the serial composition and symplifying the term, we obtain $[()\leadsto\True]$.

Although the translation of Boolean blocks is rather involved, the result obtained 
especially after basic Isabelle simplifications is simple and intuitive, as shown above.
Moreover, for the translation of a Boolean block we do not need to consider its context, and
the correctness of the translation can be assessed locally. Basically an element $e$ in a 
conjunction $(e\land\ldots)$ 
is replaced by $((e\not=0)\land\ldots)$. By creating the class $\mathsf{numeral\_nzero}$
and the instantiations to $\mathsf{bool}$ and $\mathsf{real}$, the typing of $e$ ($e:\mathsf{bool}$ 
or $e:\mathsf{real}$, $\ldots$) defines the semantics of the expression $e\not=0$.

\section{Generic Translations}
\label{sec:generic}

The approach described so far works well for diagrams that do not mix
values of different types (Boolean, real) in operations. However,
there are some diagrams that are accepted by Simulink and cannot
be translated with the approach described above due to type mismatch.
Fig.~\ref{figsimulinkaccepts} and \ref{figsimulinkrejects} give three examples of this kind of diagrams.

\hide{
\begin{figure}
\subfloat[Adding a Boolean and a real.]{\begin{tikzpicture}
\node[draw, minimum width = 4ex, minimum height = 4ex](a){$2:\bool$};
\node[draw, minimum width = 4ex, minimum height = 4ex, below = 1ex of a](b){$3:\real$};
\node[fit = (a) (b)](d){};
\node[draw, minimum width = 4ex, minimum height = 8ex, right = 3ex of d](aa){$\Add$};
\draw[-latex'](a.east) -- (a -| aa.west);
\draw[-latex'](b.east) -- (b -| aa.west);
\draw[-latex'](aa.east) -- ++(3ex,0);
\end{tikzpicture}
\label{fig:bool-real-add}
}

\subfloat[Boolean into integral.]{\begin{tikzpicture}
\node[draw, minimum width = 4ex, minimum height = 4ex](a){$2:\bool$};
\node[draw,minimum width = 5ex, minimum height = 4ex, right = 3ex of a](inte){$\int$};
\draw[-latex'](a) -- (inte);
\draw[-latex'](inte.east) -- ++(3ex,0);
\end{tikzpicture}
\label{fig:boo-inte}
}

\subfloat[Adding Boolean and real and integrating the result.]{\begin{tikzpicture}
\node[draw, minimum width = 4ex, minimum height = 4ex](a){$2:\bool$};
\node[draw, minimum width = 4ex, minimum height = 4ex, below = 1ex of a](b){$3:\real$};
\node[fit = (a) (b)](d){};
\node[draw, minimum width = 4ex, minimum height = 8ex, right = 3ex of d](aa){$\Add$};
\draw[-latex'](a.east) -- (a -| aa.west);
\draw[-latex'](b.east) -- (b -| aa.west);

\node[draw,minimum width = 5ex, minimum height = 4ex, right = 3ex of aa](inte){$\int$};
\draw[-latex'](aa) -- (inte);
\draw[-latex'](inte.east) -- ++(3ex,0);

\end{tikzpicture}
\label{fig:bool-real-add-inte}
}
\caption{Conversions.}
\label{fig:ill-types}
\end{figure}
}

Fig.~\ref{figsimulinkaccepts} illustrates diagrams accepted by Simulink, while the diagram represented
in Fig.~\ref{figsimulinkrejects} is not accepted by Simulink. The simulation
of leftmost diagram from Fig.~\ref{figsimulinkaccepts} gives $4$ ($2:\bool$
results in $\True$, and then converted to real is $1$). The rightmost diagram
from Fig.~\ref{figsimulinkaccepts} is equivalent to a diagram
where constant $4$ is input for an integral block. However none of
these diagrams result in correct translations when using the method presented
so far. This is due to type mismatches: \rs
\[
\begin{array}{l}
(\const(2:\bool)\para\const(3:\real))\circ\Add\\
(((\const(2:\bool)\para\const(3:\real))\circ\Add)\para\Id)\circ[(x:\real),s\leadsto s,s+x\cdot\mathit{dt}]\\
(\const(2:\bool)\para\Id)\circ[(x:\real),s\leadsto s,s+x\cdot\mathit{dt}]
\end{array} \rd
\]
In the first case, we try to add a Boolean to a real. The second example contains the first example
as a sub-diagram, and it has the same type incompatibility. In the third
example the output of $\const(2:\bool)$ of type $\bool$ is used
as the input for the first component of $[(x:\real),s\leadsto s,s+x\cdot\mathit{dt}]$
which expects a real.

To be able to translate these diagrams, we use type variables
instead of the concrete types $\bool$, $\real$, $\ldots$. Because
we work with expressions containing arithmetic and Boolean operations,
we need to use type variables of appropriate classes. For example,
to translate the leftmost diagram from Fig.~\ref{figsimulinkaccepts}, we cannot
just use an arbitrary type $\tp a$ because $\tp a$ must be of class $\num$
for the constants $2$ and $3$, and of class $\plus$. In fact only
class $\num$ is required here because $\plus$ is a subclass of $\num$.
The generic translation of this diagram is: \rs
\[
\begin{array}{ll}
\mathsf{ConstA}(x:\tp a:\num)  =\const(2:\tp a),\ \
\mathsf{ConstB}(y:\tp a:\num)  =\const(3:\tp a)\\[0.5ex]
A(x,y) =(\mathsf{ConstA}(x)\para\mathsf{ConstB}(y))\circ[a,b\leadsto a+b]
\end{array} \rd
\]
In this translation, we only need to specify the types for the constants
as discussed in Section \ref{sec:constants}. However, when we use
the type variable $\tp a$ for numeric constants $1,2,\ldots$, then
we must specify it using the class $\num$. If the expression involving
the elements of type $\tp a$ contains some other operators, then we must
include also the classes defining these operators. For example we
need to have: $\mathsf{ConstA}(x:\tp a:\{\num,\mult\})=\const((2:\tp a)\cdot3)$.
To simplify this we introduce a new class $\mathsf{simulink}$ that
contains all mathematical and Boolean operators as well as all real
functions that can occur in Simulink diagrams. \rs
\[
\begin{array}{l}
\mathsf{class\ simulink=zero+numeral+minus+uminus+power+ord\ +}\\
\qquad\mathsf{fixes\ s\_exp,\ s\_sin:\tp a\to\tp a}\ \ | \ \
\mathsf{fixes\ s\_and:\tp a\to\tp a\to\tp a}\\
\qquad\ldots\\
\qquad\mathsf{assume\ numeral\_nzero[simp]:\ }(\forall a.\mathsf{numeral}(a)\not=0)
\end{array} \rd
\]
Class $\zero$ contains the symbol $0$, class $\num$ contains the
numbers $1,2,\ldots$, classes $\mathsf{minus}$ and $\mathsf{uminus}$
contains the binary and unary minus operators, class $\mathsf{power}$ contains
the power and multiplication operators, and class $\mathsf{ord}$
contains the order operators. Because the real functions $\mathsf{exp}$,
$\mathsf{sin}$, $\ldots$ and the Boolean functions are defined just
for reals and Boolean types respectively, and they do not have generic type classes,
we introduce the generic versions of these functions and operators
in the class $\mathsf{simulink}$ ($\mathsf{s\_exp,s\_sin,\ldots,s\_and,\ldots}$).
Additionally we assume that constant $0$ is different from all numeric
constants $1,2,\ldots$. 

Using this new class the translation of the rightmost diagram from Fig.~\ref{figsimulinkaccepts} 
is given by \rs 
\[
\begin{array}{l}
\mathsf{ConstA}(x:\tp a:\simu)=\const(\mathsf{s\_bool}(2:\tp a))\\
\mathsf{ConstB}(y:\tp a:\simu)=\const(3:\tp a)\\
\mathsf{Integral}(\mathit{dt}:\tp a:\simu)=[s,x\leadsto s,s+x\cdot dt]\\
\Add=[(x:\tp a:\simu),y\leadsto x+y]\\
A(x,y,\mathit{dt})=(((\mathsf{ConstA}(x)\para\mathsf{ConstB}(y))\circ\Add)\para\Id)\circ\mathsf{Integral}(\mathit{dt})
\end{array} \rd
\]
The inferred type of $A$ is $A(x:\tp a:\simu,y:\tp a,dt:\tp a):\tp a\ptto\tp a\times\tp a$

In this generic translation there are some details to consider when translating
a constant block of type Boolean like the ones from Fig.~\ref{figsimulinkaccepts}
($2:\bool$). In order to use $A(x,y,dt)$ in the end, we still need
to instantiate the type variable $\tp a$. In this case, it would be
appropriate to instantiate $\tp a$ with type $\real$. If we simply
use $\const(2:\tp a)$ in definition of $\mathsf{ConstA}$, then when
instantiating $\tp a$, we will obtain the constant $2$ and we will
add it to $3$ resulting in $5$, and this is not the result
obtained when simulating the diagram in Simulink. To preserve the
Simulink semantics in the generic case, we translate Boolean constants
using a function $\mathsf{s\_bool}$ which for a parameter $x$
returns $1$ if $x$ is different from $0$ and $0$ otherwise:
\[
\mathsf{\qquad definition\ s\_bool}(x):=\mathsf{if}\ x\not=0\ \mathsf{then}\ 1\ \mathsf{else}\ 0
\]
The typing of $x:\tp a$ and of $\mathsf{s\_bool}(x):\tp b$ defines again
a more precise semantics for $\mathsf{s\_bool}(x)$. For example if both $\tp a$ and $\tp b$
are $\mathsf{bool}$, then $\mathsf{s\_bool}(x) = x$.
\ifdefined\techrep \else Similarly, we define instantiations for $\bool$ and $\real$ for all the generic
functions defined in the $\mathsf{simulink}$ class. These instantiations are detailed in \cite{preoteasa:dragomir:tripakis:2016}. \fi

\ifdefined\techrep
Assume now that we have an expression with type variables of class $\simu$ \rs
\[
(x:\tp a:\simu)+\mathsf{s\_exp}(y)=2.5\land\mathsf{s\_and}(u: \tp b:\simu,v)=1 \rd
\]
When we instantiate type variables $\tp a$ and $\tp b$ with the types
$\real$ and $\bool$ respectively we want to obtain the expression:
$
x+\mathsf{exp}(y)=0.5\land u\land v 
$.
For this we need to introduce type instantiations for the $\simu$ class.
For $\real$ have: \rs
\[
\begin{array}{l}
\mathsf{instantiation\ \real:simulink}=\\
\qquad\mathsf{s\_exp}(x:\real):=\mathsf{exp}(x)\\
\qquad\mathsf{s\_and}(x:\real,y:\real):=\mathsf{if}\ x\not=0\land y\not =0\ \mathsf{then}\ 1\ \mathsf{else}\ 0\\
\qquad\ldots
\end{array} \rd
\]
The proper definitions for the arithmetic operations are
already provided in the Isabelle library. Similarly we need to provide
the instantiation for $\bool$: \rs
\[
\begin{array}{l}
\mathsf{instantiation\ \bool:simulink}=\\
\qquad(0:\bool):=\False \ \ | \ \ 
(\num(a):\bool):=\True\\
\qquad\mathsf{s\_and}(x:\bool,y:\bool):=x\land y\\
\qquad\ldots
\end{array} \rd
\]
In the instantiation to type $\real$ we have chosen to implement
the Boolean operations using real numbers. This definition is motivated by expressions
that mix Boolean and real values as in the examples from Fig.~\ref{figsimulinkaccepts}.
This formalization is fair, but also sufficient, because the Boolean values can be easily embedded within the
real numbers. However, when we have an addition block $\Add:(\tp a:\simu)\times \tp a \ptto \tp a$
in the generic translation it does not make sense to instantiate $\tp a$
to $\bool$. If we do so accidentally, at least we do not provide
definitions for the arithmetic operations within the Boolean domain.
This will prevent any simplifications to occur for the wrongly interpreted
expressions. However, we must provide definitions for the numeric
constants $1,2,\ldots$, otherwise we cannot prove the assumption
$\mathsf{numeral\_nzero}$ for the Boolean instantiation. If we choose
not to use the assumption $\mathsf{numeral\_nzero}$ in $\simu$ then
we do not need these definitions, except for $1$ that plays the role
of $\True$.
\fi

We implemented this strategy in our Simulink to Isabelle model translator under the 
$\mathsf{-generic}$ option. When this option is missing, then all blocks
are defined using their specific types. If this option is given, then
only type variables of class $\simu$ are used. 

Additionally, we implemented the option $\mathsf{-type}\ \mathit{isabelle\_type}$
with an Isabelle type parameter, which adds a new definition
where it instantiates all type variables to the type parameter.

For example, if we apply the translation using the options $\mathsf{-const}$, $\mathsf{-generic}$,
and $\mathsf{-type\ real}$
to the rightmost diagram from Fig.~\ref{figsimulinkaccepts}, we obtain: \rs
\[
\begin{array}{l}
\mathsf{ConstA}(x:\tp a:\simu):=\const(\mathsf{s\_bool}(2:\tp a))\\
\mathsf{ConstB}(y:\tp a:\simu):=\const(3:\tp a)\\
\mathsf{Integral}(\mathit{dt}:\tp a:\simu):=[s,x\leadsto s,s+x\cdot dt]\\
\Add:=[(x:\tp a:\simu),y\leadsto x+y]\\
\mathsf{A}(x,y,\mathit{dt}):=(((\mathsf{ConstA}(x)\para\mathsf{ConstB}(y))\circ\Add)\para\Id)\circ\mathsf{Integral}(\mathit{dt})\\
\mathsf{A\_type}(\mathit{dt}):=\mathsf{A}(0:\real,0:\real,\mathit{dt}:\real)
\end{array} \rd
\]
and also the simplified versions $\mathsf{A}$ and $\mathsf{A\_type}$:\rs 
\[
\begin{array}{lll}
\mathsf{A}(x,y,\mathit{dt})=[s\leadsto s,s+(\mathsf{1+3})\cdot dt] & \mbox{ and } & \mathsf{A\_type}(\mathit{dt})=[s\leadsto s,s+4\cdot dt]
\end{array} \rd
\]
In the generic version $\mathsf{s\_bool}(2)$ is automatically simplified
to $1$ using the definition of $\mathsf{s\_bool}$ and the assumption
$\mathsf{numeral\_nzero}$, and in $\mathsf{A\_type}$ the expression 
$1+3$ is further simplified to $4$. In $\mathsf{A\_type}$ we can eliminate
the variables providing types for constants because these types are now
instantiated to $\real$.

\section{Implementation and Correctness}
\label{sec:impl}

The mechanism presented above for translating Simulink diagrams
is implemented in the {Refinement Calculus of Reactive Systems} framework, available from \url{http://rcrs.cs.aalto.fi}. 
In this framework, Simulink diagrams are translated into Isabelle theories, where diagrams are modeled using predicate transformers. 
The framework allows to perform various analysis on the formal model such
as simplification, compatibility checking, safety property verification and
simulation.

In order to handle a large set of diagrams, we introduced three translation options: $\mathsf{-const}, 
 \mathsf{-generic}$, and $\mathsf{-type}\ \mathit{isabelle\_type}$, where each solves different possible corner cases.
These options allow some control over the translation process.
\ifdefined\techrep \else
More details about these options are available in the
extended version of this work \cite{preoteasa:dragomir:tripakis:2016}.
\fi

\ifdefined\techrep
The option
$\mathsf{-const}$ must be used when the diagram contains constants with types that
are not reflected in the overall type of the diagram. In general these
diagrams can be seen as pathological, and they can be re-factored to eliminate this
problem. The option $\mathsf{-const}$ can be used for all diagrams, but when this issue 
is not present, it generates additional parameters that
are not needed. Because well designed diagrams do not have this problem,
we designed the translation such that the $\mathsf{-const}$ option is disabled by default.

The option $\mathsf{-generic}$ is needed in diagrams that mix Boolean and
numeric values. When using this option, the typing of the translation is more general than the typing of the original diagram, as we use type variables everywhere.
However, if some Boolean values are used in numerical calculations
(like in Fig.~\ref{figsimulinkaccepts}), then the same type variable
is used for the Boolean values as well as for the numerical expression.
The idea is that this type variable is instantiated  
to the numerical type (by the user) and all Boolean operations are implemented using numbers. 
This is the only situation when a type of the original diagram (Boolean)
is translated into a different (numeric) type. We can also see this case as pathological, and the user can decide to add explicit conversions
in the diagram to avoid this problem.

Finally, the option $\mathsf{-type}\ \mathit{isabelle\_type}$ does not influence the
translation itself, but it can be used to easily 
instantiate all type variables in the translation to a given Isabelle type.
This is the only option that may produce a more concrete translation, compared to
the original Simulink diagram.
\fi

In most cases the translation results in representations that are
more general (with respect to typing) than the original diagram.
The only exception is when Boolean values in numeric expressions are used, 
as noted \ifdefined\techrep above and \fi in Section \ref{sec:generic}.
Instantiating the remaining type variables can be done such that the types of the translation match the types inferred by Simulink. Obtaining
a more general typing ensures the correctness of this approach.
When Boolean constants and operations are implemented
using numeric constants and operations, the correctness is also
ensured because this implementation is quite straightforward
and standard.

We have extensively tested all combinations of interactions of numeric
and Boolean blocks, and we carefully implemented the observed behavior.
We have also tested this technique on several case studies, 
as well as an industrial example:
the Fuel Control System (FCS) benchmark from Toyota \cite{JinDKUB14}.
We validated our approach by simulation. 
We translated the Isabelle systems into
Python and we simulated them, resulting in the same behavior 
(modulo a small simulation error) as the Simulink simulation.
\ifdefined\techrep
All examples have been translated using all combinations of the  
options. In each case the results obtained were as expected:
either a type mismatch or the same values as those provided by Simulink simulation.
For the FCS case study, when translating it only with the $\mathsf{-const}$ option,
we have detected a type mismatch which was not signaled by Simulink.
An easy correction is to add a conversion block. Even though Simulink,
as well as other programming languages, implicitly perform this conversion we believe that having a fully correct type checked model 
will increase the confidence of the generated code. 

\delete{
The validation of the obtained results was performed through simulation. Our framework
allows to generate from Isabelle a Python simulation script corresponding to the model
under evaluation. We have simulated the models in the Python script and in Simulink, and we have compared 
the obtained values. The differences between our approach and Simulink are small -- $10^{-5}$ in absolute 
value for the FCS case study --, which can be related to the numerical integration method, simulation time step,
representation of floats in memory, etc. Therefore, the simulation plots are nearly identical. 
A more detailed discussion about the validation of the translation in general can be found in \cite{DBLP:conf/spin/DragomirPT16}.
}
\else
A more detailed discussion about the validation of the translation in general and the results obtained on the FCS case study
can be found in \cite{DBLP:conf/spin/DragomirPT16,preoteasa:dragomir:tripakis:2016}.
\fi

\section{Conclusions}
\label{sec:concl}
In this work we presented a type inference technique for Simulink diagrams
using the type inference of Isabelle. 
\ifdefined\techrep
With the exception of calculating the arities
of the wires in a Simulink diagram, this
\else This \fi technique is treating the basic blocks
of the diagram locally, without knowledge of their context.

Although the translation process is quite involved with many cases,
the result of the translation (after basic Isabelle simplifications)
is simple and meaningful.

The results presented in this paper are implemented in a framework for translating
Simulink diagrams into theories for the  Isabelle theorem prover, available 
from \url{http://rcrs.cs.aalto.fi/}.

\bibliographystyle{abbrv}
\bibliography{bibl}

\end{document}